\documentclass[%
 reprint,
 amsmath,amssymb,
 aps,
]{revtex4-1}

\usepackage{graphicx}
\usepackage{dcolumn}
\usepackage{bm}
\usepackage{multirow}

\begin{document}

\preprint{APS/123-QED}

\title{Chiral symmetries and Majorana fermions in coupled magnetic atomic chains on a superconductor}

\author{Jinpeng Xiao}
 \altaffiliation{}
\author{Jin An}%
\affiliation{
National Laboratory of Solid State Microstructures and Department of Physics, Nanjing University, Nanjing 210093, China
}


\date{\today}

\begin{abstract}
We study the magnetic structures and their connections with topological superconductivity due to proximity effect for coupled magnetic atomic chains deposited on a superconductor. Several magnetic phases are self-consistently determined, including both the coplanar and non-coplanar ones. For a $N$-chain triangular atomic ladder, topologically nontrivial superconducting states can always be realized, but strongly depend on its magnetic structure and the number of atomic chains. When $N$ is even, the topologically nontrivial states with noncoplanar structures are characterized by $\mathbb{Z}_{2}$ invariants, while the topologically nontrivial noncoplanar states with an odd $N$ are characterized by integer $\mathbb{Z}$ invariants, due to the presence of a new chiral symmetry. The new chiral symmetry for the noncoplanar states is found to be robust against the on-site disorder, as long as the crystal reflection symmetry is respected.

\begin{description}
\item[PACS numbers]{71.10.Pm, 73.20.-r, 73.63.Nm, 74.45.+c}
\end{description}
\end{abstract}
\maketitle

\section{\label{sec:level1}Introduction}

Topological superconductors (TSCs) have drawn much attention in condensed matter physics over the last decade, since they may host at defects unpaired Majorana fermions \cite{Kiteav2001}, which obey non-Abelian statistics\cite{Kiteav2003,Stern2004} and thus may have potential applications in topological quantum computing\cite{Read2000,Ivanov2001,Das2006,Stone2006,Nayak2008}.
Many approaches have been considered to creating TSCs and their neutral excitations, including proximity effects between topological insulators and superconductors\cite{Fu2008,Williams2012}, semiconductors with strong spin-orbit couplings (SOCs)\cite{Sau2010,Lutchyn2010}, doped topological insulators\cite{Hor2010,Wray2011,Pavan2011,Hung2013}, and quantum-dot-superconductor arrays\cite{Sau2012}. A common feature of all these approaches is that SOC plays an important role in creating Majorana bound states (MBSs)\cite{Elliott2015}.

A new way to explore TSCs and MBSs has been developed recently by depositing magnetic atoms on a conventional superconductor\cite{Heimes2014,Nadj2014,Li2014,Heimes2015,Dumitrescu2015,Brydon2015,Ebisu2015,Ojanen2015,Peng2015,Sau2015,Choy2011,Martin2012,Pientka2013,Vazifeh2013,Bernd2013,Yazdani2013,Jelena2013,Nakosai2013,Kim2014,Poyhonen2014,Falko2014,Reis2014,Rontynen2014,Weststrom2015,Sedlmayr2015,Hu2015}. Topologically nontrivial superconducting states can be induced in the absence of SOC if the magnetic atoms may be arranged into perfect helical structures \cite{Choy2011,Martin2012,Pientka2013,Vazifeh2013,Bernd2013,Yazdani2013,Jelena2013,Nakosai2013,Kim2014,Poyhonen2014,Falko2014,Reis2014,Rontynen2014,Weststrom2015,Sedlmayr2015,Hu2015}. Experimentally, array of magnetic atoms and their spins can be manipulated accurately by STM\cite{Serrate2010,Khajetoorians2011,Khajetoorians2012}. Topological superconductivity in ferromagnetic chains as long as 50nm on the surface of a $\mathrm{Pb}$ superconductor has recently been realized\cite{Nadj2014}. On the other hand, the magnetic atoms on the top of a superconductor can be regarded as magnetic impurities in a 2D superconductor, thus resulting in the formation of the Yu-Shiba-Rusinov midgap bound states\cite{Yu1965,Shiba1968,Rusinov1969}. These localized bound states become extended and form a band when their wave functions overlap. The effective RKKY interactions between the localized atoms' magnetic moments via the exchange with itinerant electrons may thus favor a helical texture\cite{Bernd2013,Vazifeh2013,Jelena2013,Kim2014,Reis2014,Hu2015}, leading to a topological superconducting state.

In this paper, we investigate the self-organized magnetic textures for coupled chains of magnetic atoms on the surface of a superconductor and their connections with the topologically nontrivial phases. Our purpose is to find out what kind of uniform magnetic structures can coexist with the induced superconductivity and which ones can be topologically nontrivial and how they are classified according to their symmetries. It is found that for coupled triangular $N$-chain magnetic adatoms, more complex magnetic structures than perfect helix can be self-organized, which also support the topological superconductivity. The noncoplanar magnetic phases even have a new chiral symmetry, which plays an important role in classifying the topological states. The chiral symmetry is characterized by a crystal reflection symmetry which is present when the chain number $N$ is odd but absent when $N$ is even.

The paper is organized as follows. The effective model and its Bogoliubov-de Gennes (BdG) Hamiltonian for the magnetic adatoms placed on a s-wave superconductor is introduced and discussed in section II. The magnetic structures are self-consistently determined and the topological phases are classified for the $N$-chain triangular ladders in section III. Different chiral symmetries are introduced for the coplanar and
noncoplanar phases with odd chain number. Extension to coupled square chains are also made. Finally, we summarize our results and briefly discuss the connections with experiments in section IV.

\section{\label{sec:level2}The Effective Model}
We consider a triangular ladder of magnetic atoms deposited on a s-wave superconductor. Superconductivity is induced in the magnetic atoms due to proximity effect. When the degrees of freedom of the substrate superconductor have been integrated out\cite{Vazifeh2013}, the hybrid system can then be thought of as an effective one for the magnetic atoms. The Hamiltonian can be given by,
\begin{equation}
\begin{split}
\mathcal{H}=&-\sum_{<ij>}t_{ij}c_{i}^{\dag}c_{j}-\mu\sum_{i}c_{i}^{\dag}c_{i}
-J\sum_{i}c_{i}^{\dag}\mathbf{S}_{i}\cdot\bm{\sigma}c_{i}\\
&+\sum_{i}(\Delta c_{i\uparrow}^{\dag}c_{i\downarrow}^{\dag}+H.c.),
\end{split}
\end{equation}
where $t_{ij}$ is the itinerant electrons' hopping amplitude between the nearest neighboring (NN) sites on the triangular ladder. Here $c_{i}^{\dag}=(c_{i\uparrow}^{\dag},c_{i\downarrow}^{\dag})$, with $c_{i\sigma}^{\dag}$ the electron creation operators on site $i$. $\mathbf{S}_{i}$ is the classical magnetic moment of the magnetic atom, $J$ represents its ferromagnetic coupling strength to itinerant electrons, and $\bm{\sigma}$=($\sigma_{x},\sigma_{y},\sigma_{z}$) is the vector of spin Pauli matrices. The last term is the pairing term due to proximity effect, and $\Delta$ the induced pairing parameter which is assumed to be uniform.

In Nambu basis $(c_{i\uparrow},c_{i\downarrow},c_{i\downarrow}^{\dag},-c_{i\uparrow}^{\dag})^{T}$,
the corresponding BdG equations can be written as $H_{ij}\psi_{nj}=E_{n}\psi_{ni}$, $\psi_{ni}=(u_{ni\uparrow},u_{ni\downarrow},v_{ni\downarrow},-v_{ni\uparrow})^{T}$, with $E_{n}\geq0$ the excitation energy and
\begin{equation}
\begin{split}
H_{ij}=-t_{ij}\tau_{z}-\delta_{ij}(J\mathbf{S}_{i}\cdot\bm{\sigma}+\mu\tau_{z}-\Delta\tau_{x}).
\end{split}
\end{equation}
Here $\tau_{i}$ are Pauli matrices operating in the particle-hole space. The BdG Hamiltonian $H_{ij}$ has a natural particle-hole $C$ symmetry: $CHC^{-1}=-H$, where $C=\tau_{y}\sigma_{y}K$, with $K$ the complex conjugation. Thus the quasi-1D coupled chains can be characterized by a $\mathbb{Z}_{2}$ invariant due to topological classification\cite{Andreas2008,Ryu2010}.
\begin{figure}
\scalebox{1.0}{\includegraphics[width=0.45\textwidth]{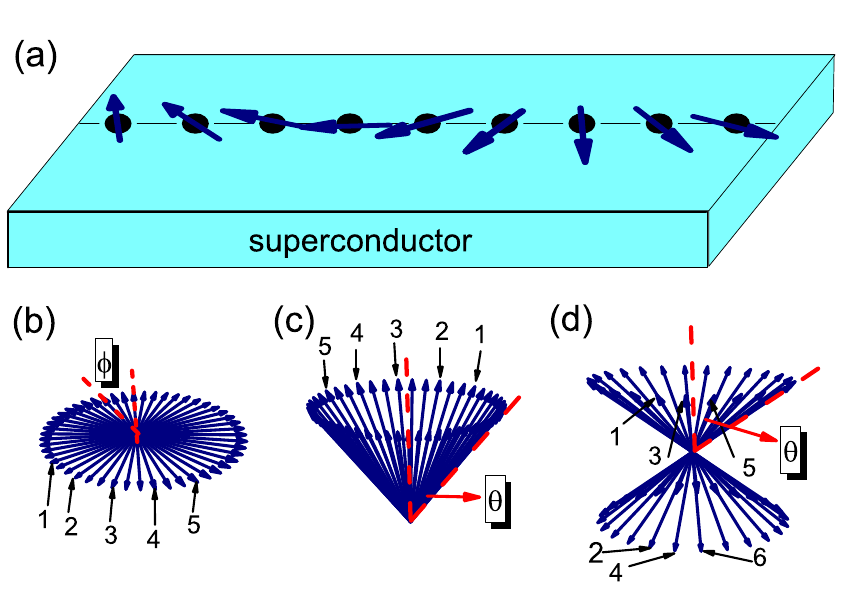}}
\caption{\label{fig:epsart} (a)Schematic depiction of the arrangement of magnetic atoms with a helical magnetic structure on a s-wave superconductor. In (b)-(d) the self-consistently determined magnetic structures for each magnetic atomic chain in a ladder are schematically shown, where (b) is a perfect helix, (c) a single-cone helix and (d) a double-cone helix.}
\end{figure}

Starting from a random configuration of $\mathbf{S}_{i}$, the BdG equations are numerically solved and then electron spin momentum $\mathbf{s}_{i}=\frac{1}{2}\langle c_{i\alpha}^{\dag}\bm{\sigma}_{\alpha\beta}c_{i\beta}\rangle$ is determined, which is given by $\mathbf{s}_{i}=\frac{1}{2}\sum_{n}[u_{ni\alpha}^{\ast}\bm{\sigma}_{\alpha\beta}u_{ni\beta}f_{F}(E_{n})+v_{ni\alpha}\bm{\sigma}_{\alpha\beta}v_{ni\beta}^{\ast}f_{F}(-E_{n})]$, with $f_{F}(x)=\frac{1}{1+e^{x/\mathrm{k_{B}T}}}$ the Fermi distribution function. In the next iteration, the classical magnetic moment $\mathbf{S}_{i}$ is directed along $\mathbf{s}_{i}$, thus in this way the BdG equations for a $N$-chain ladder system can be solved iteratively by typically performing hundreds of iterations. To extract the low-energy physics, temperature is set to be $\mathrm{T}=0$. In the following, energy is measured in unit of the NN hopping amplitude $t$, and the order parameter $\Delta$ is fixed to be $\Delta=0.15$ unless otherwise stated.

\section{\label{sec:level3} MAGNETIC STRUCTURES AND TOPOLOGICALLY NONTRIVIAL STATES}
For a generic $N$-chain triangular ladder, we have self-consistently solved the BdG equations and determined the classical magnetic moments $\mathbf{S}_{j}$ and found five noncollinear magnetic phases in total. A common feature for all the phases is that each atomic chain can take only the following three structures, regardless of $N$, and the magnetic structures for coupled chains can then be regarded as combinations of the three by symmetries. The three one-chain structures are depicted schematically in figure 1. In spherical coordinates, $\mathbf{S}_{j}$ can be expressed as
$\mathbf{S}_{j}=S(\sin(\theta_{j})\cos(j\phi),\sin(\theta_{j})\sin(j\phi),\cos(\theta_{j}))$, with $\theta_{j}$, $j\phi$ the polar and azimuth angles of $\mathbf{S}_{j}$ respectively.

When projected on a plane, each of the three structures becomes a perfect helix with $\phi$ the pitch angle, as seen in the figure. In addition, the latter two noncoplanar ones satisfy the following conditions: $\theta_{i}\equiv\theta$ is a constant for a single-cone helix(see figure 1(c)) while $\theta_{2i+1}=\pi-\theta_{2i}\equiv\theta$ for a double-cone helix(see figure 1(d)), i.e., along each atomic chain, atoms' magnetic moments alternatively belong to different cones for a double-cone helix. Here $\theta$ is the cone angle. The perfect helix can also be viewed as a special case of single- or double-cone ones with $\theta\equiv\pi/2$. Though the cone angle $\theta$ may vary from chain to chain, the pitch angle $\phi$ is identical for all chains. This is interpreted as the spin instability of system at $q=2\phi$. The conical structures for some parameter regions are believed to be self-organized to reduce energy of system by lifting degeneracies of the band crossings of the helical structures.

\subsection{Single-chain and double-chain ladders}
We start our discussions from the simplest case of the single atomic chain. For a single-chain system, the only self-consistent(SC) noncollinear magnetic phase is a coplanar one with a perfect helical structure(figure 2(a)). This state is spatial uniform, which can be seen by introducing the local spin axis defined by $\mathbf{S}_{j}$ and then the corresponding annihilating operators at each site by performing spin rotations  $c_{j\uparrow}=e^{-i(j\phi/2)}d_{j\uparrow}$, $c_{j\downarrow}=e^{i(j\phi/2)}d_{j\downarrow}$. Only two terms in Hamiltonian (1) are potentially affected by this transformation. The hopping term becomes spin-dependent and thus plays the role of the effective SOC in the system. The spin exchange $J$ term becomes spatial uniform since all local magnetic moments are now directed along the same direction in the local spin axes. Therefore the single-chain Hamiltonian in $k$ space can be expressed as $\mathcal{H}=\frac{1}{2}\sum_{k}\psi_{k}^{\dag}H_{k}^{0}\psi_{k}$, where $\psi_{k}^{\dag}$ is Nambu basis $\psi_{k}^{\dag}=(d_{k\uparrow}^{\dag},d_{k\downarrow}^{\dag},d_{-k\downarrow},-d_{-k\uparrow})$. $H_{k}^{0}$ can be given as,
\begin{equation}
H_{k}^{0}=(\xi_{k,\phi}-\mu)\tau_{z}+\eta_{k,\phi}\tau_{z}\sigma_{z}+h\sigma_{x}+\Delta\tau_{x}.
\end{equation}
Here $h=JS$, $\xi_{k,\phi}=-2\mathrm{cos}\frac{\phi}{2}\mathrm{cos}k$, $\eta_{k,\phi}=-2\mathrm{sin}\frac{\phi}{2}\mathrm{sin}k$. The topological property of a 1D superconducting system can be characterized by a $\mathbb{Z}_2$-valued invariant, which is topologically protected by particle-hole symmetry, satisfying $CH_{-k}^{0}C^{-1}=-H_{k}^{0}$. This $\mathbb{Z}_{2}$ value can be expressed as the signature $M$ of the Pfaffian of an antisymmetric matrix, which is that of the Hamiltonian in its Majorana representation\cite{Kiteav2001,Yazdani2013}. The single magnetic atomic chain on a superconductor is a well studied topic, and it is now known that a perfect helical texture for atoms with proximity induced conventional superconductivity may enable the system to be a TSC\cite{Choy2011,Martin2012,Pientka2013,Vazifeh2013,Bernd2013,Yazdani2013,Jelena2013,Kim2014,Falko2014,Reis2014,Rontynen2014,Weststrom2015,Hu2015}. The signature $M$ for a fixed helical structure has been studied before\cite{Vazifeh2013,Yazdani2013} and can be given analytically as $M=\mathrm{sgn}\{[\delta^{2}-(\xi+\mu)^{2}][\delta^{2}-(\xi-\mu)^{2}]\}$, where $\delta^{2}=h^{2}-\Delta^{2}$ and $\xi=-2\cos\frac{\phi}{2}$. When $M=-1$, the system is topologically nontrivial. We illustrate $M$ in figure 2(b)(both green and yellow colors represent $M=-1$), together with the SC result of our calculations, indicating that most of the data are located at the topologically nontrivial region. This result is consistent with that of the minimization of the free energy in Refs. \cite{Vazifeh2013,Reis2014,Hu2015}, and the susceptibility calculated in Refs. \cite{Bernd2013,Jelena2013,Kim2014}. Generally, the change of $M$ is accompanied with gap closing and reopening at $k=0$ or $\pi$.
\begin{figure}
\scalebox{1.0}{\includegraphics[width=0.45\textwidth]{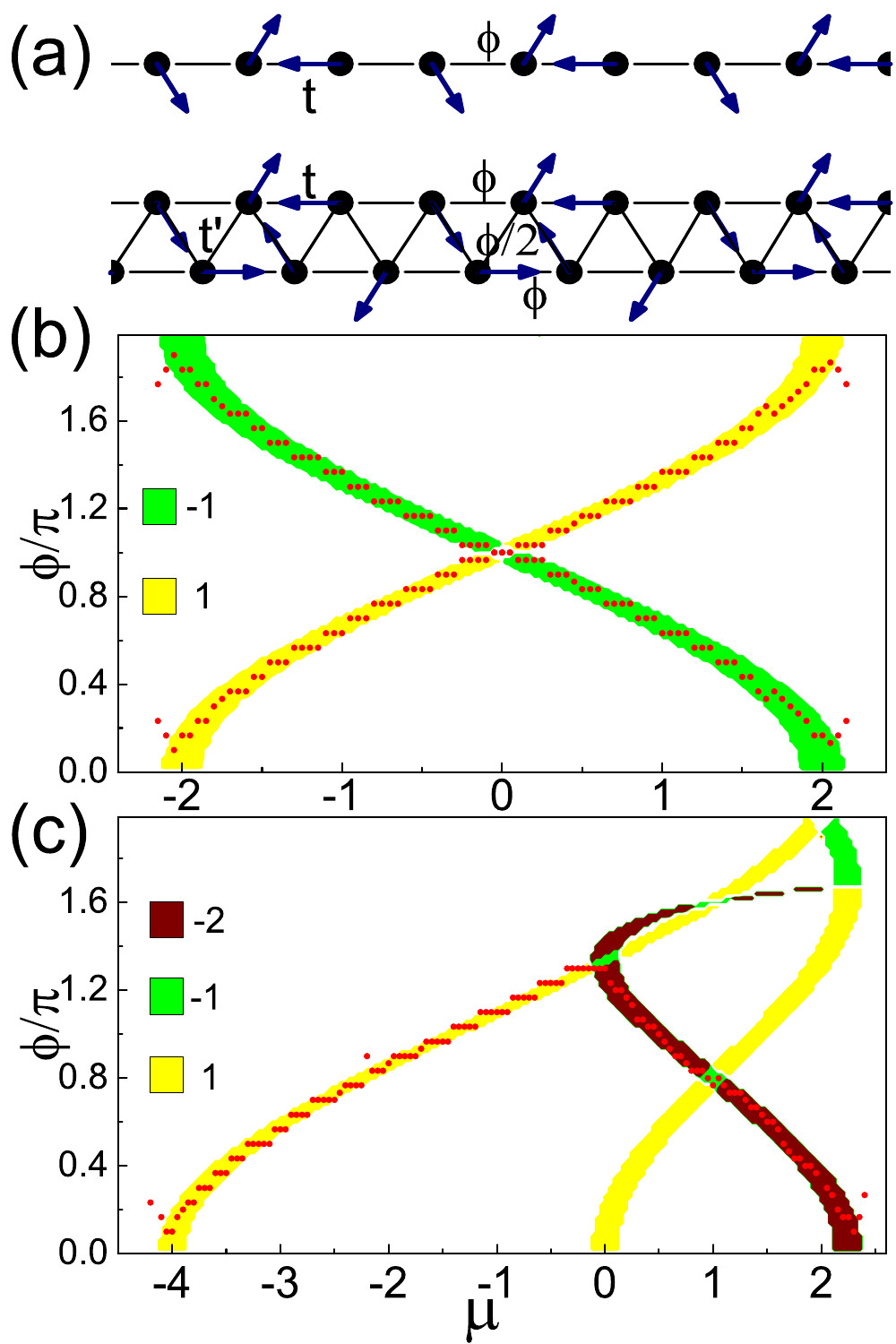}}
\caption{\label{fig:epsart} (a)Schematic helical magnetic structures of the magnetic atoms self-consistently determined for a single and coupled double chains. The single-chain and double-chain topological phase diagrams as functions of the pitch angle $\phi$ and the chemical potential $\mu$ are depicted respectively in (b) and (c), where the green or yellow colored regions denote the ones with the nontrivial $\mathbb{Z}_{2}$ values protected by the particle-hole symmetry, while different colors(including the brown region in(c)) represent different nontrivial $\mathbb{Z}$ values protected by the chiral symmetry. The red solid dots are the self-consistent results for systems with sizes $1\times60$ and $2\times60$ respectively. Here $JS=0.2$, $t'=1$.}
\end{figure}

For a coupled double chain with the interchain hopping amplitude $t'$, so long as $t'$ is not small enough, our SC result suggests that the ground state can be viewed as a zigzag single chain which also has a perfect helical structure\cite{Yazdani2013}, as seen schematically in figure 2(a). Actually along the zigzag direction this `superchain' contains atoms of the two single chains alternatively, so $t'$ and $t$ can be regarded as the hopping amplitudes between the NN and the next NN neighbors of the `superchain'. Compared with the single-chain case, the signature $M$ can be given accordingly, $M=\mathrm{sgn}\{[\delta^{2}-(\xi+\xi'-\mu)^{2}][\delta^{2}-(\xi-\xi'-\mu)^{2}]\}$, with $\xi'=-2t'\cos(\frac{\phi}{4})$. In figure 2(c) the regions with nontrivial $M=-1$ are colored by green or yellow. When $t'<t$, the SC results show that there also exists a small doping region where each single chain is ferromagnetic while it is antiferromagnetically coupled to its neighboring chain. This collinear magnetic state is by itself obviously topologically trivial, but can show topological nontrivial features when a supercurrent is flowing through the system\cite{Heimes2014}.

For both single-chain and double-chain cases, there exists a chiral symmetry $\{H_{k},S\}=0$, with $S=\tau_{y}\sigma_{z}$\cite{Poyhonen2014}. The existence of the chiral symmetry and its combination with the $C$ symmetry lead to $T$ symmetry: $TH_{-k}T^{-1}=H_{k}$, with $T=\sigma_{x}K$. Thus the system belongs to BDI class due to topological classification\cite{Andreas2008,Ryu2010}. Any 1D chiral symmetric system of BDI class with translational symmetry is characterized by a $\mathbb{Z}$-valued invariant $\nu$ which can be expressed in terms of the chiral operator $S$ and the Hamiltonian $H_{k}$ satisfying $H_{k+2\pi}=H_{k}$ as follows\cite{ayrynen2011},
\begin{equation}
\begin{split}
\nu=\frac{1}{4\pi i}\int_{\mathrm{BZ}}dk\mathrm{Tr}[SH_{k}^{-1}\partial_{k}H_{k}].
\end{split}
\end{equation}
The chiral symmetric system with nonzero integer $\nu$ is topologically nontrivial. Physically, the absolute value $|\nu|$ indicates the number of MBSs at each end protected by the chiral symmetry. This can be interpreted in a simple way. Denote the zero-energy Majorana fermions at each end as $\gamma_{i}$, i=1,2,..$|\nu|$. The only possible terms which can open energy gaps take the forms of $i\gamma_{i}\gamma_{j}$.
\begin{figure}
\scalebox{1.0}{\includegraphics[width=0.45\textwidth]{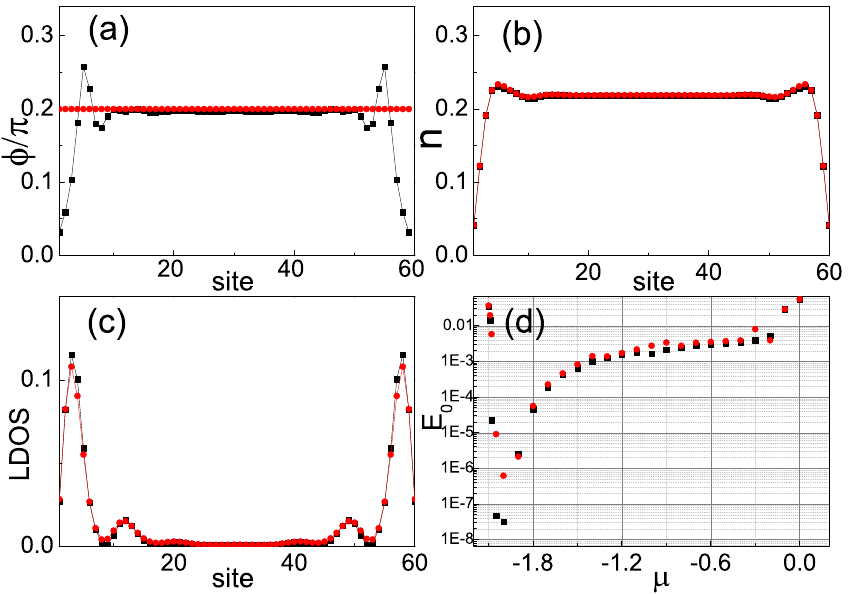}}
\caption{\label{fig:epsart} Comparison of features between the self-consistent and non-self-consistent results for a singe-chain system with $60$ lattice sites under open boundary conditions. (a)The angles between the nearest magnetic moments. (b)Electron density distributions. (c)The spatial distributions for the lowest energy state $E_{0}$. (d)The lowest energy $E_{0}$ versus the chemical potential $\mu$. Here $\mu=-1.9$ for (a)-(c) and $JS=0.2$. The solid squares(circles) correspond to the (non-)self-consistent results.}
\end{figure}
Under $T$ symmetry operation with $T^{2}=1$, since each $\gamma_{i}$ is invariant, these terms change signs and so are forbidden. Thus the system remains gapless and the number of MBSs at each end is robust. Straightforward calculations show that the parameter areas of the nonzero $\nu$ always completely cover that of the nontrivial $\mathbb{Z}_{2}$, at which $\nu$ takes odd integers. Furthermore, an additional new area with $\nu=-2$ appears for the double-chain case, as depicted in figure 2(c). The SC result for the double chain is also shown in figure 2(c), indicating that most of the data are located in the two topological branches with $\nu=1$ and $\nu=-2$ respectively. Starting from a topologically trivial state and varying a parameter adiabatically will lead to quantum phase transitions to topologically nontrivial ones. When the parameter is tuned to enter the $\nu=\pm1$ region, gap will close and reopen in $k$ space at $k=0$ or $k=\pi$, while when tuned to the $\nu=-2$ region, the gap closing and reopening points will occur at another asymmetric $k$ points $\pm k_{c}$ determined by $\mathrm{cos}k_{c}=-(\frac{4t}{t'}\cos(\frac{\phi}{4}))^{-1}$.

To exhibit MBSs and study the boundary effects on them, we impose open boundary conditions for a single-chain system and self-consistently determine the boundary magnetic moments $\mathbf{S}_{i}$. As expected, the direction of the magnetic moment $\mathbf{S}_{i}$ is greatly distorted. Although $\mathbf{S}_{i}$ is deviated obviously from the perfect helix at both ends, they are still located at the same plane, forming a planar structure and thus preserving the chiral symmetry. The angles between the neighboring $\mathbf{S}_{i}$ are shown in figure 3(a). Interestingly, those boundary effects almost hardly affect the MBS' lowest energy and its spatial local density distribution, compared with those of the non-SC result which is simply performed by imposing open boundary without taking any distortion of $\mathbf{S}_{i}$ into account. The insensitivity of the open boundary conditions on the MBSs implies the properties of these localized states still mainly depend on the bulk properties, not the boundary effects.
\begin{figure}
\scalebox{1.0}{\includegraphics[width=0.45\textwidth]{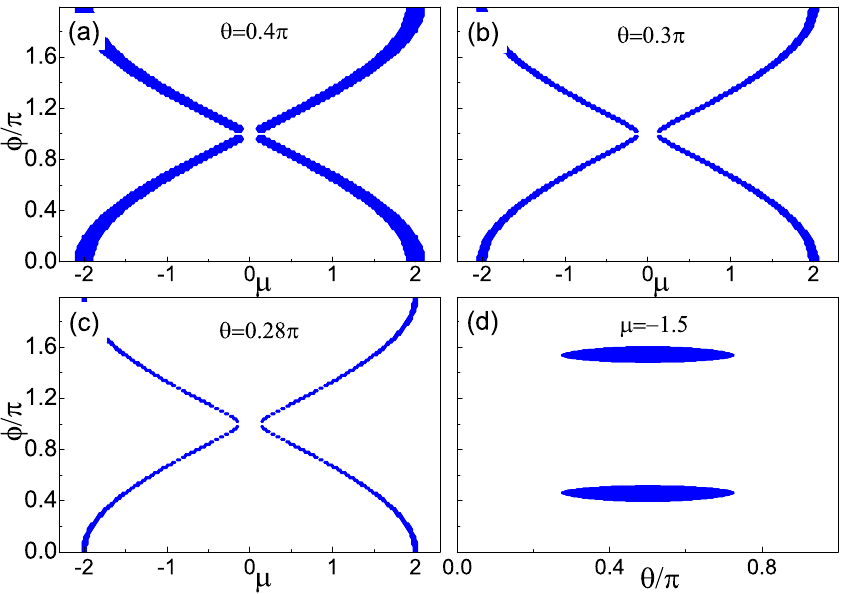}}
\caption{\label{fig:epsart} (a)-(c)Topological phase diagram as a function of the pitch angle $\phi$ and the chemical potential $\mu$ for a single chain with a double-cone magnetic structure, where the cone angle $\theta=0.4\pi$ for (a), $\theta=0.3\pi$ for (b), $\theta=0.28\pi$ for (c) respectively. (d)Topological phase diagram as a function of $\phi$ and $\theta$ for fixed $\mu=-1.5$. Here the colored area denotes the topological states with a nontrivial $\mathbb{Z}_{2}$ value.}
\end{figure}

Although the single-cone and double-cone structures as illustrated in figure 1(c),(d) are not found in our SC calculations for a single chain, we would like to study their topological properties since they are intrinsically connected with the $N$-chain topological states. For the single-cone case, the signature $M$ shows similar behavior to that for the planar helix, as shown in figure 2(b), except that the bulk gap will close when the cone angle $\theta$ is increased up to a critical value. For the double-cone case, the situation is more interesting. The signature $M$ takes the following analytic form,
\begin{equation}
\begin{split}
M=&\mathrm{sgn}\{[\delta^{2}-(\xi+\mu)^{2}][(\delta^{2}-(\xi-\mu)^{2}]\\
&+4h^{2}\xi^{2}\cos^{2}\theta\},
\end{split}
\end{equation}
which demonstrates that the topological region shrinks as the cone angle $\theta$ decreases from $\pi/2$, as shown in figure 4. When $\theta$ becomes smaller than a critical value $\theta_{c}=\mathrm{cos}^{-1}(\sqrt{1-\Delta^{2}/h^{2}})$, the topological nontrivial region vanishes. The chiral symmetry does not exist for the single-cone and double-cone single-chain states, so they are only characterized by the $\mathbb{Z}_{2}$-valued invariants.

\subsection{$N$-chain ladders($3\leq$N$\leq$6)}

\subsubsection{Self-consistently determined magnetic structures.}

The SC results for a coupled triangular $N$-chain ladder with even $N$ show that it is possible for its magnetic structure to be collinear. These collinear states can be characterized by the uniform ferromagnetism in each chain but without bulk magnetization as a whole. Since in the absence of SOC, it is impossible for a collinear state to be topologically nontrivial, we only focus on the noncollinear magnetic states. The SC results exhibit various structures for $N>2$, including five noncollinear magnetic phases, which can be regarded as the combinations of the three basic ones displayed in figure 1(b)-(d). By using the unit vectors $\mathbf{a}=a(1,0)$ and $\mathbf{b}=a(-\frac{1}{2},-\frac{\sqrt{3}}{2})$ with $a$ the lattice constant, each lattice site of the ladder can be represented as $j\mathbf{a}+l\mathbf{b}$, where $l=0,1,...N-1$ is the chain index and $j$ the site index for each chain. Thus the five magnetic phases can be divided into the following three categories:
\begin{enumerate}
\item Coplanar phase I(CI), $S_{j}(l)=S(\cos[(j-l/2)\phi],\sin[(j-l/2)\phi],0)$ and Coplanar phase II(CII), $S_{j}(l)=S(-1)^{s_{l}}(\cos[(j-l/2)\phi],\sin[(j-l/2)\phi],0)$, where $s_{l}=[\frac{l}{2}]$ if $N$ is even, otherwise $s_{l}=[\frac{l}{2}]$ if $l\leq\frac{N}{2}$ and $s_{l}=s_{N-1-l}$ if $l>\frac{N}{2}$. Here $[x]$ is the integer part of $x$;
\item Single-cone phase(S), $S_{j}(l)=S(\sin\theta_{l}\cos[(j-l/2)\phi],\sin\theta_{l}\sin[(j-l/2)\phi],\cos\theta_{l})$, where $\theta_{l}=\pi-\theta_{N-1-l}$;
\item Double-cone phase I(DI) and II(DII) for odd $N$, $S_{j}(l)=S(\sin\theta_{l}\cos[(j-l/2)\phi],\sin\theta_{l}\sin[(j-l/2)\phi],(-1)^{j}\cos\theta_{l})$, where $\theta_{l}=\pi-\theta_{N-1-l}$, and $\theta_{l}=\frac{\pi}{2}$ if $l$ is even(odd) for DI(DII).
\end{enumerate}
\begin{figure}
\scalebox{1.0}{\includegraphics[width=0.45\textwidth]{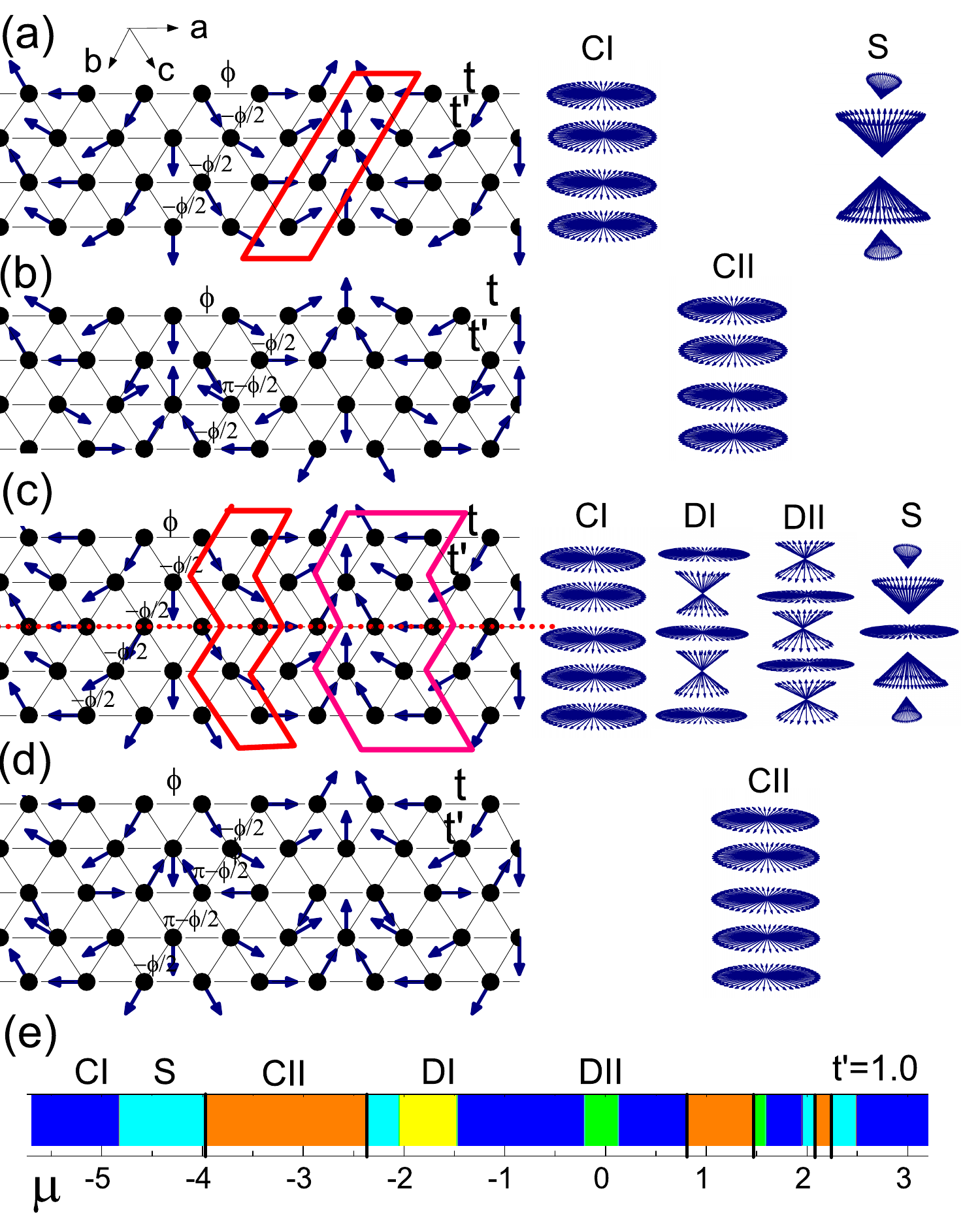}}
\caption{\label{fig:epsart} Schematic depiction of the self-consistently determined uniform noncollinear magnetic structures for the four-chain cases (a),(b), and the five-chain cases (c),(d). Each chain can take only one of the three basic structures: the planar helix, the single-cone helix, or the double-cone helix. The closed loops in (c) denote the unit cells chosen for the single-cone or double-cone states and the red doted line given in (c) denotes the reflection line for the chiral symmetric states. (e)The self-consistently determined doping phase diagram for coupled five chains with $t'=1$, $JS=0.2$, where the black lines denote the discontinuous first-order transitions.}
\end{figure}
These magnetic phases and the doping phase diagram for $N=5$ are shown in figure 5. The doping phase diagram indicates that most of the magnetic structures can be competing in several different doping ranges, which is consistent with the multiple-chain band structure and the topological phase diagram introduced below. All these phases are spatially uniform, which can be understood by introducing as before the local spin-up and spin-down annihilating operators $d_{i\sigma}(l)$.

\subsubsection{Chiral symmetries, topological stability and topological classification.}

Using the unit cell shown in figure 5(a) and in the basis of $\Psi_{k}^{\dag}=(\psi_{k}^{\dag}(0),\psi_{k}^{\dag}(1),\cdots,\psi_{k}^{\dag}(N-1))$ with $\psi_{k}^{\dag}(l)=(d_{k\uparrow}^{\dag}(l),d_{k\downarrow}^{\dag}(l),d_{-k\downarrow}(l),-d_{-k\uparrow}(l))$, the Hamiltonian for the planar CI phase can be written as,
\begin{equation}
\begin{split}
&H_{k}^{CI}=H_{k}^{0}+H_{k}^{1}\\
&H_{k}^{1}=t'(\cos\frac{k}{2}\Gamma_{x}-\sin\frac{k}{2}\Gamma_{y})\tau_{z}\\
&(\xi_{k/2,\phi/2}+\eta_{k/2,\phi/2}\sigma_{z}),
\end{split}
\end{equation}
where $H_{k}^{0}$ is the single-chain Hamiltonian given by Eq. (3), and $H_{k}^{1}$ the interchain hopping term. Here
\begin{equation}
\begin{array}{c c}
\Gamma_{x}=
\left(
\begin{array}{ccccc}
  0&   1&   &   &   \\
  1&   0&   1&   &  \\
  &   1&   0&   1&  \\
  &   &  \cdot& \cdot& \cdot \\
  &   &   &   1&   0  \\
\end{array}
\right),
\Gamma_{y}=
\left(
\begin{array}{ccccc}
  0&   -i&   &   &   \\
  i&   0&   -i&   &  \\
  &   i&   0&   -i&  \\
  &   &  \cdot& \cdot& \cdot \\
  &   &   &   i&   0  \\
\end{array}
\right),
\end{array}
\end{equation}
are two $N\times N$ hermitian matrices acting on the interchain index space. For the planar CI phase the signature $M$ of the Pfaffian can be given analytically ,
\begin{equation}
\begin{split}
M=&\mathrm{sgn}\{[\delta^{2}-(2\cos\frac{\phi}{2}-\mu)^{2}]^{s}\prod_{n=1}^{N}(\delta^{2}-\xi_{n}^{2})\},
\end{split}
\end{equation}
where $\xi_{n}=-2t\cos\frac{\phi}{2}-4t'\cos\frac{\phi}{4}\cos(\frac{n\pi}{N+1})-\mu$, and $s=1(0)$ if $N$ is odd(even). The factors in the expression of $M$ actually form a polynomial of $\mu$ of degree $2N$ (or $2N+2$) if $N$ is even(odd), thus generically there exist $N$ (or $N+1$) branches in topological phase diagram $\phi$ versus $\mu$ if $N$ is even(odd), as shown in figure 6. The fact that topological region occurs alternatively as a function of $\mu$ can be understood easily in the weak pairing limit where $\Delta$ is much smaller than all the other relevant energy scales, since the Fermi points of the underling normal system for the half Brillouin zone has to be odd to make the system to be topologically nontrivial\cite{Kiteav2001}. Since $S=\tau_{y}\sigma_{z}$ anticommutes with $H_{k}^{1}$, the planar CI phase has the chiral symmetry: $\{H_{k}^{CI},S\}=0$. It can be easily shown that any planar magnetic phase shares the same chiral symmetry\cite{Poyhonen2014}. Thus $\{H_{k}^{CII},S\}=0$ and so both CI and CII phases belong to the BDI class which can be characterized by the $\mathbb{Z}$-valued invariants. In figure 6 both the $\mathbb{Z}_{2}$ invariant values $M$ and the $\mathbb{Z}$ invariant values $\nu$ are shown for the planar phases. As a comparison, the SC results are depicted in figure 6($\mathrm{a}_{2}$)-($\mathrm{e}_{2}$), which indicates that due to RKKY interactions, the system can be easily self-organized to be topologically nontrivial. Similar to the double-chain planar states, the region with the nontrivial $\mathbb{Z}$ values $\nu$ completely covers that with the nontrivial $\mathbb{Z}_{2}$ ones. At these `old' branches $\nu$ takes odd integer values $\pm1$. Additional `new' topologically nontrivial branches appear with $\nu=\pm2$, and at the edges of these new branches, the gap opening or closing points generically occur at the non-time-reversal-invariant $k$ points($k\neq0,$ or $\pi$). When the `new' and `old' branches overlap, regions with $\nu=\pm3$ may occur, as shown in figure 6 by the blue or orange colored regions. As long as $JS>\Delta$, topological region with nontrivial $M$ is possible except the planar CII phase for even $N$ for which $JS$ has a critical value of 0.34 when $\Delta=0.15$, $N=4$. Only when $JS>0.34$, there exist nontrivial $M$ regions, as illustrated by figure 6($\mathrm{d}_{3}$).
\begin{figure*}
\centering
\scalebox{1.0}{\includegraphics[width=0.95\textwidth]{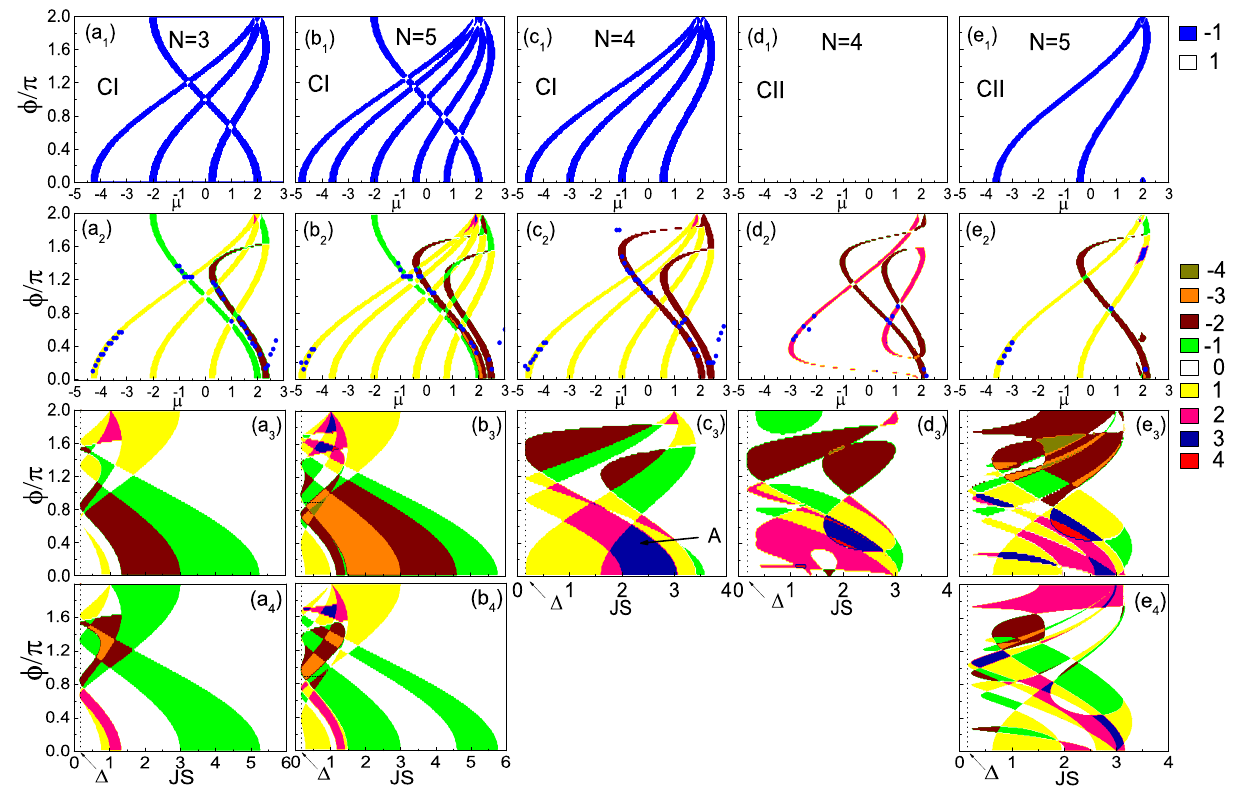}}
\caption{\label{fig:epsart} Topological phase diagrams for the coplanar magnetic phases. As functions of the pitch angle $\phi$ and the chemical potential $\mu$, the $\mathbb{Z}_{2}$ invariant values are shown in ($\mathrm{a}_{1}$)-($\mathrm{e}_{1}$), while the corresponding $\mathbb{Z}$ invariant values $\nu$ are shown in ($\mathrm{a}_{2}$)-($\mathrm{e}_{2}$), for fixed $JS=0.2$, where the solid circles are the self-consistent results. In the lower two rows, the $\mathbb{Z}$ invariant values as functions of $\phi$ and $JS$ for fixed $\mu=1$ are shown in ($\mathrm{a}_{3}$)-($\mathrm{b}_{3}$), ($\mathrm{a}_{4}$)-($\mathrm{b}_{4}$), while the others are for fixed $\mu=-1$. Here $t'=0.8$. The chain number $N=3$ for ($\mathrm{a}_{1}$), ($\mathrm{a}_{2}$), ($\mathrm{a}_{3}$) and ($\mathrm{a}_{4}$), $N=4$ for ($\mathrm{c}_{1}$), ($\mathrm{c}_{2}$), ($\mathrm{c}_{3}$), ($\mathrm{d}_{1}$), ($\mathrm{d}_{2}$), ($\mathrm{d}_{3}$), and $N=5$ for the others. Here ($\mathrm{a}_{2}$)-($\mathrm{e}_{2}$) and ($\mathrm{a}_{3}$)-($\mathrm{e}_{3}$) correspond to the chiral symmetry relevant to $\tau_{y}\sigma_{z}$, while ($\mathrm{a}_{4}$),($\mathrm{b}_{4}$) and ($\mathrm{e}_{4}$) to that relevant to $\Gamma\tau_{y}\sigma_{z}$. }
\end{figure*}

The Hamiltonian $H_{k}^{S}$ for the single-cone S states is exactly the same to $H_{k}^{CI}$, except that the $h$ term in $H_{k}^{CI}$ is replaced accordingly by $h\Gamma_{z}^{1}\sigma_{x}+h\Gamma_{z}^{2}\sigma_{z}$ in $H_{k}^{S}$, where $\Gamma_{z}^{1}$, $\Gamma_{z}^{2}$ are the diagonal matrices acting on the interchain index space, with $\Gamma_{z}^{1}=\mathrm{diag}\{\sin\theta_{0},\sin\theta_{1},\cdots,\sin\theta_{N-1}\}$, and $\Gamma_{z}^{2}=\mathrm{diag}\{\cos\theta_{0},\cos\theta_{1},\cdots,\cos\theta_{N-1}\}$. It can be easily checked $H_{k}^{S}$ has no chiral symmetry. However, if the chain number $N$ is odd and a zigzag unit cell as displayed in figure 5(c) is chosen, the Hamiltonian has a chiral symmetry. This new Hamiltonian takes a slightly different form $H_{k}^{S'}$, which differers from $H_{k}^{S}$ only in that $\Gamma_{y}$ matrix is replaced by $\widetilde{\Gamma}_{y}$. Together with $\widetilde{\Gamma}_{y}$, we also introduce another $N\times N$ hermitian $\Gamma$ matrix, which are given by,
\begin{equation}
\begin{array}{c c}
\widetilde{\Gamma}_{y}=
\left(
\begin{array}{ccccc}
  0&   -i&   &   &   \\
  i&   0&   i&   &  \\
  &   -i&   0&   -i&  \\
  &   &  \cdot& \cdot& \cdot \\
  &   &   &   i(-1)^{N}&   0  \\
\end{array}
\right),
\Gamma=
\left(
\begin{array}{ccccc}
  &   &   &   &1   \\
  &   &   &1   &  \\
  &   &1   &   &  \\
  &\cdot  & & &\\
  1  &   &   &  &  \\
\end{array}
\right).

\end{array}
\end{equation}
It can be easily checked that $[\Gamma,\Gamma_{x}]=\{\Gamma,\Gamma_{y}\}=0$, while $\Gamma$ and $\widetilde{\Gamma}_{y}$ anticommute $\{\Gamma,\widetilde{\Gamma}_{y}\}=0$ if $N$ is even, but commute $[\Gamma,\widetilde{\Gamma}_{y}]=0$ if $N$ is odd. Meanwhile, for any matrices of $\Gamma_{z}^{1}$, $\Gamma_{z}^{2}$ types satisfying $\theta_{l}+\theta_{N-1-l}=\pi$, we have $[\Gamma,\Gamma_{z}^{1}]=\{\Gamma,\Gamma_{z}^{2}\}=0$. Therefore, if $N$ is odd, $H_{k}^{S'}$ has a chiral S symmetry $\{H_{k}^{S'},S\}=0$ with $S=\Gamma\tau_{y}\sigma_{z}$. If $N$ is even, $H_{k}^{S'}$ remains chiral symmetry breaking. The $\Gamma$ matrix can be viewed as the reflection operator in the interchain index space which has a reflection line along the middle chain(see figure 5(c)), and transforms each chain to its reflection symmetric partner. We note that only when the unit cell respects the crystal reflection symmetry for odd $N$, the Hamiltonian in $k$ space can be chiral symmetric. The nonexistence of the chiral symmetry of the Hamiltonian $H_{k}^{S}$ for even $N$ is due to the nonexistence of the reflection line, while for odd $N$, the nonexistence can be viewed as a result of improperly choosing a unit cell without a reflection line.

From this viewpoint, the planar phases actually have two chiral operators, if the reflection symmetric zigzag unit cell is chosen. For CI phase, the other chiral operator is $S'=\Gamma\tau_{y}\sigma_{z}$ since $[H_{k}^{CI},\Gamma]=0$ in this representation. For CII phase, the other chiral operator $S'$ takes the similar form but with a more complicated $N$-dependent minor diagonal $\Gamma$ matrix ($\Gamma=\left(\begin{smallmatrix}  &  & & & 1\\ &  & & 1& \\  &  & -1& & \\  & 1 & & & \\1 &  & & &  \end{smallmatrix}\right)$ for $N=5$ as an example). Accordingly, there is another $\mathbb{Z}$ topological invariant $\nu'$ relevant to $S'$ for the coplanar phases. We compare $\nu$ with $\nu'$ for the exactly same parameters in figure 6($\mathrm{a}_{3}$)-($\mathrm{a}_{4}$), figure 6($\mathrm{b}_{3}$)-($\mathrm{b}_{4}$) and figure 6($\mathrm{e}_{3}$)-($\mathrm{e}_{4}$). Both $\nu$ and $\nu'$ share exact the same phase boundaries. In each topological region, the difference between the values of $\nu$ and $\nu'$ is an even integer. There exist topological regions with $\nu'\neq0$ but $\nu=0$ which are protected by the chiral symmetry relevant to $S'$ but not by that to $S$. Since $S'$ is characterized by the reflection operator $\Gamma$, for the system with a zigzag ends, a local perturbation breaking the reflection symmetry will break the chiral symmetry and can thus remove the MBSs at chain ends protected only by $S'$, but leave the MBSs protected by $S$ unchanged.

In the basis of $(\Psi_{kA}^{\dag},\Psi_{kB}^{\dag})$ and using the reflection symmetric unit cell as shown in figure 5(c), the Hamiltonian for the double-cone D phases can be written as,
\begin{equation}
H_{k}=\left(
                  \begin{array}{cc}
                    h_{k+} & e^{-ik/2}t_{k} \\
                    e^{ik/2}t_{k} & h_{k-} \\
                  \end{array}
                \right),
\end{equation}
where the block-diagonal terms $h_{k\pm}$ are given as $h_{k\pm}=-\mu\tau_{z}+\Delta\tau_{x}+h\Gamma_{z}^{1}\sigma_{x}\pm h\Gamma_{z}^{2}\sigma_{z}+\xi'\Gamma_{x}\tau_{z}-\eta'\widetilde{\Gamma}_{y}\tau_{z}\sigma_{z}$ while the hermitian $t_{k}$ is $t_{k}=\tau_{z}(\xi_{k/2,\phi}+\eta_{k/2,\phi}\sigma_{z})+\frac{t'}{2}(\xi_{k/2,\phi/2}\Gamma_{x}-\xi'_{k/2,\phi/2}\widetilde{\Gamma}_{y})\tau_{z}+\frac{t'}{2}(\eta_{k/2,\phi/2}\Gamma_{x}+\eta'_{k/2,\phi/2}\widetilde{\Gamma}_{y})\tau_{z}\sigma_{z}$, with $\xi'=-t'\cos\frac{\phi}{4}$, $\eta'=-t'\sin\frac{\phi}{4}$, $\xi'_{k,\phi}=-2\cos\frac{\phi}{2}\sin k$, $\eta'_{k,\phi}=-2\sin\frac{\phi}{2}\cos k$. In a similar way, one can check that only if $N$ is odd, DI and DII phases have the chiral symmetry $\{H_{k}^{D},S\}=0$ with the same $S$ operator to that of the S phases(for DII phases, the chiral symmetry is preserved only when the middle chain's cone angle $\theta_{(N-1)/2}=\frac{\pi}{2}$).

Although $\Gamma$ is a reflection operator appearing in the chiral operator $S$, reflecting the crystal symmetry of the ladder, $\Gamma$ itself does not commute with the Hamiltonian: $[H_{k}^{S(D)},\Gamma]\neq0$ due to the presence of the magnetic order. This indicates the Hamiltonian $H_{k}^{S(D)}$ can not be reduced by any crystal symmetry operator. The proximity induced order parameter $\Delta$ is assumed here to be uniform, which is actually not necessary for the preservation of the chiral symmetry. Even if the order parameter takes different values $\Delta_{l}$ for different chains but preserves the reflection symmetry: $\Delta_{l}=\Delta_{N-1-l}$, which is a more realistic situation, the $\Delta$ term in $H_{k}^{S(D)}$ should be replaced by one proportional to $\Gamma_{z}^{1}\tau_{x}$. The chiral symmetry is preserved for all the above mentioned phases since $\{\Gamma_{z}^{1}\tau_{x},S\}=0$. Similarly, it is very likely that each chain has a different site energy but respects the reflection symmetry, thus the chemical potential $\mu$ term in $H_{k}^{S(D)}$ has to be replaced by one proportional to $\Gamma_{z}^{1}\tau_{z}$. This kind of perturbation still cannot break the chiral symmetry since $\{\Gamma_{z}^{1}\tau_{z},S\}=0$.
\begin{figure*}
\scalebox{1.0}{\includegraphics[width=0.95\textwidth]{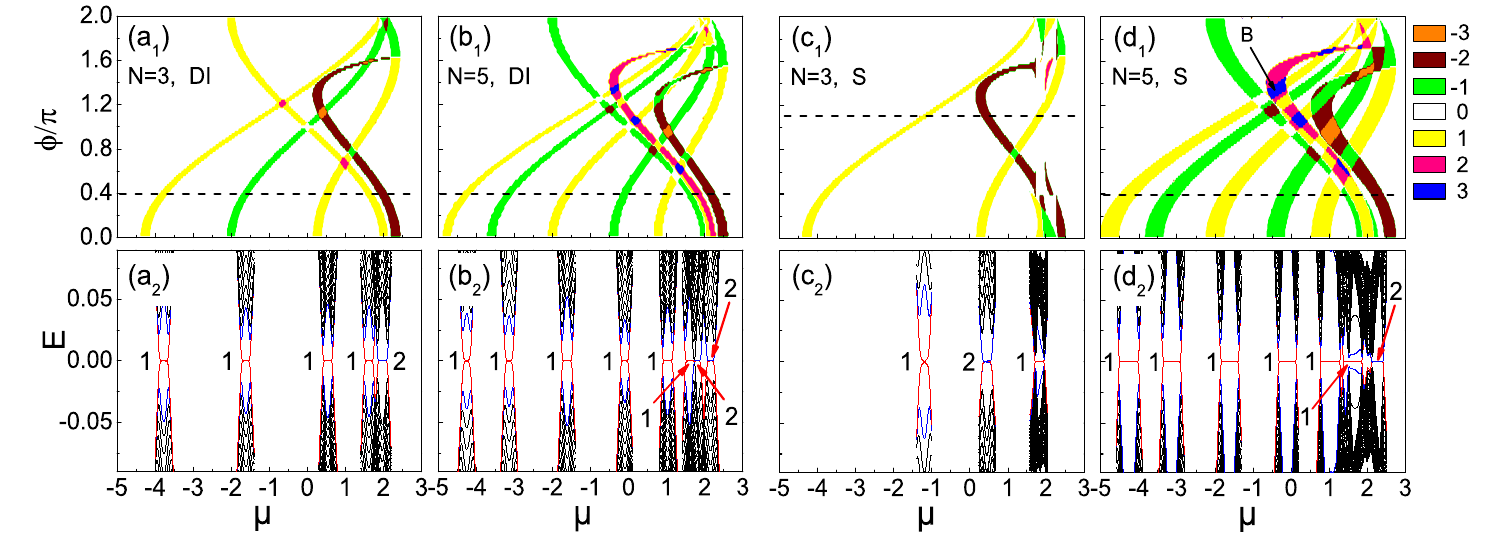}}
\caption{\label{fig:epsart} Topological phase diagrams and the low-lying energy spectra for the non-coplanar magnetic phases with the chiral symmetry. The left panels are for the double-cone DI states while the right panels are for the single-cone S states. The $\mathbb{Z}$ invariant values $\nu$ are shown in ($\mathrm{a}_{1}$)-($\mathrm{d}_{1}$), while the corresponding open-boundary energy spectra for a fixed pitch angle as denoted by the dotted lines are depicted in ($\mathrm{a}_{2}$)-($\mathrm{d}_{2}$) respectively. The parameters chosen are: $N=3$, $\theta_{0}=\pi-\theta_{2}=0.4\pi,\theta_{1}=0.5\pi$, $JS=0.2$ for ($\mathrm{a}_{1}$),($\mathrm{a}_{2}$); $N=5$, $\theta_{1}=\pi-\theta_{3}=0.4\pi,\theta_{0}=\theta_{2}=\theta_{4}=0.5\pi$, $JS=0.2$ for ($\mathrm{b}_{1}$),($\mathrm{b}_{2}$); $N=3$, $\theta_{0}=\pi-\theta_{2}=0.08\pi,\theta_{1}=0.5\pi$, $JS=0.32$ for ($\mathrm{c}_{1}$),($\mathrm{c}_{2}$); and $N=5$, $\theta_{0}=\pi-\theta_{4}=0.35\pi,\theta_{1}=\pi-\theta_{3}=0.4\pi,\theta_{2}=0.5\pi$, $JS=0.32$ for ($\mathrm{d}_{1}$),($\mathrm{d}_{2}$). The integer numbers in the spectra denote the number of the pairs of the Majorana zero-energy bands. }
\end{figure*}

\begin{figure}
\centering
\scalebox{1.0}{\includegraphics[width=0.45\textwidth]{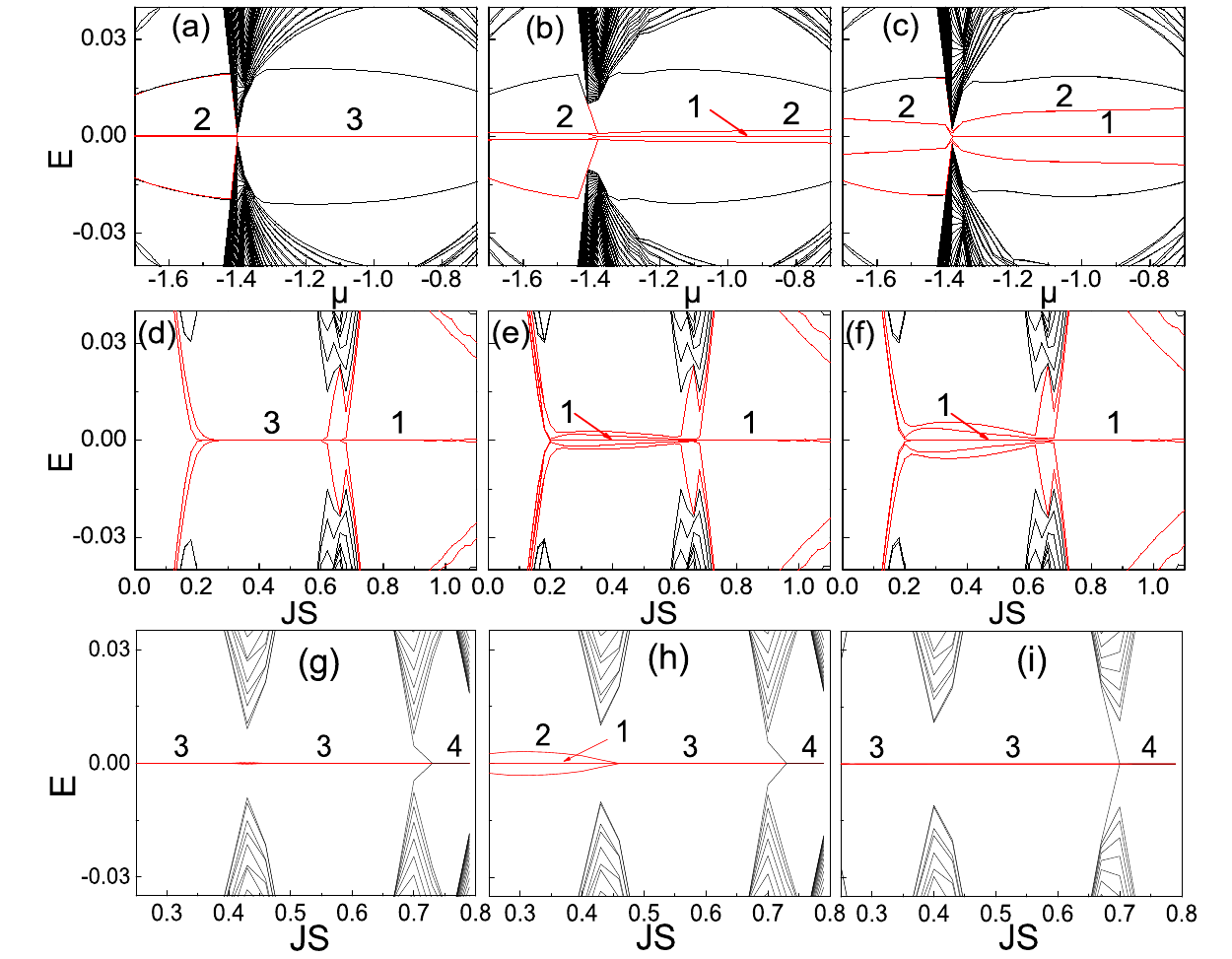}}
\caption{\label{fig:epsart} The open-boundary low-lying energy spectra exhibiting the chiral-symmetry-breaking effects. Starting from a chiral symmetric planar state with $N=4$ and $\nu=3$(point A in figure 6($\mathrm{c}_{3}$)), the effects of changing gradually one pair of the cone angles $\theta_{0}=\pi-\theta_{4}$ are shown in the upper panels, where $\Delta\theta_{0}=\pi/2-\theta_{0}=0$ for (a), $\Delta\theta_{0}=0.01\pi$ for (b), and $\Delta\theta_{0}=0.05\pi$ for (c). The middle panels show the effect of an electric field perpendicular to the chains on a noncoplanar chiral symmetric state with $N=5$ and $\nu=3$(point B in figure 7($\mathrm{d}_{1}$)), where the electric potential difference $V$ between neighboring chains is $V=0$ for (d), $V=0.01$ for (e), and $V=0.02$ for (f). The lower panels show the similar electric field effect to the middle panels, but on the coplanar phase with $N=5$, $\nu=-1$ and $\nu'=-3$(doted lines in figure 6($\mathrm{b}_{3}$),($\mathrm{b}_{4}$)) where $V=0$ for (g), $V=0.02$ for (h). In (i) a reflection symmetric perturbation of varying site energy along the interchain direction is introduced, where the site energy difference between neighboring chains is $0.02$. The integer numbers in the spectra denote the number of the pairs of the Majorana zero-energy bands or degeneracies of the non-zero-energy bands.}
\end{figure}

In the above discussions, although $\Gamma$ matrix in $S$ is a spatial operator and the $\mathbb{Z}$-valued invariant $\nu$ is defined also in $k$ space by Eq.(4), the preservation of the chiral symmetry does not rely on the translational symmetry along the chain direction. In fact, as long as the crystal reflection symmetry is respected by the on-site perturbations, $S=\Gamma\tau_{y}\sigma_{z}$ anticommutes with the real-space Hamiltonians $H^{S(D)}$ for S or D states. These on-site translation-symmetry-breaking processes include site-energy $\mu$ term, pairing $\Delta$ term, and even the magnetic exchange term where the cone angles of magnetic atoms vary along the chain direction. Thus even in the presence of on-site disorder which respects the reflection symmetry, the number of MBSs at ends is robust and topologically protected as long as the disorder strength is not strong enough to close the bulk gap.

For these noncoplanar chiral symmetric states, apart from the natural particle-hole $C$ symmetry $CH_{-k}^{S(D)}C^{-1}=-H_{k}$, with $C=\tau_{y}\sigma_{y}K$, there also exists a $T$ symmetry: $TH_{-k}^{S(D)}T^{-1}=H_{k}$, with $T=\Gamma\sigma_{x}K$. Since $T^{2}=1$, the noncoplanar chiral symmetric states belong to BDI class\cite{Andreas2008,Ryu2010}. Topological phase diagrams of $\mathbb{Z}$ invariant $\nu$ for these noncoplanar phases with odd $N$ are shown in figure 7. Also shown are the corresponding low-lying energy spectra for fixed pitch angle $\phi$ under the open boundary conditions with zigzag ends. Majorana zero-energy bands occur whenever the scanning $\mu$ is crossing nontrivial regions. These zero-energy bands have degeneracies which exactly coincide with $\nu$ vlues of the regions and thus confirm the existence of the chiral symmetry respected by the noncoplanar phases.

\subsubsection{Breaking the chiral symmetry.}

The integer number $\nu$ is protected by the chiral symmetry, but can change if it is broken. Consider a chiral symmetric state with an odd $\nu$ which has a nontrivial $\mathbb{Z}_{2}$ value $M$ and then introduce a chiral-symmetry-breaking perturbation term whose strength is characterized by $\rho$. When $\rho\neq0$, the chiral symmetry is broken, but as long as $\rho$ is sufficient small, the signature $M$ should be still nontrivial and thus the parity of the number of MBSs does not change. On the other hand, for a chiral symmetric state with an even $\nu$, any small chiral-symmetry-breaking perturbation can in principle destroy the MBSs. We consider two cases of the chiral-symmetry-breaking process: One is changing gradually one pair of the cone angles of a chiral symmetric planar state with even $N$, the other is applying an electric field perpendicular to the chains to a chiral symmetric noncoplanar state. The open-boundary low-lying energy spectra exhibiting these processes are illustrated in figure 8(a)-(f), from which it can be seen clearly that two pairs of the zero-energy bands quickly separate and deviate from the zero energy upon the decrease of the cone angle from $\pi/2$ or increase of the electric field strength.

For the odd-$N$ coplanar phase which has two chiral operators we also apply an perpendicular electric field to study the chiral-symmetry-breaking effect. In figure 8(g)-(h), we illustrate the effect for a $N=5$, $\nu=-1$, $\nu'=-3$ coplanar CI state. In this situation, the electric field breaks the refection symmetry with respect to the middle chain, and so breaks the chiral symmetry relevant to $S'=\Gamma\tau_{y}\sigma_{z}$ but still preserves the chiral symmetry relevant to $S=\tau_{y}\sigma_{z}$. The two MBSs at ends protected by $S'$ are then destroyed while the MBSs protected by $S$ survive. However, a similar perturbation but preserving the refection symmetry still cannot remove the MBSs at ends, as shown in figure 8(i).

In figure 9, we give the $\mathbb{Z}_{2}$ invariant for the chiral-symmetry-breaking noncoplanar states with even $N$, which belong to D class. It exhibits interesting behavior as a function of the pitch angle $\phi$ and $JS$. Several components of the topologically nontrivial regions shrink and then become disconnected and finally vanish upon the increase of one of the cone angles with others fixed.

\begin{figure}
\scalebox{1.0}{\includegraphics[width=0.45\textwidth]{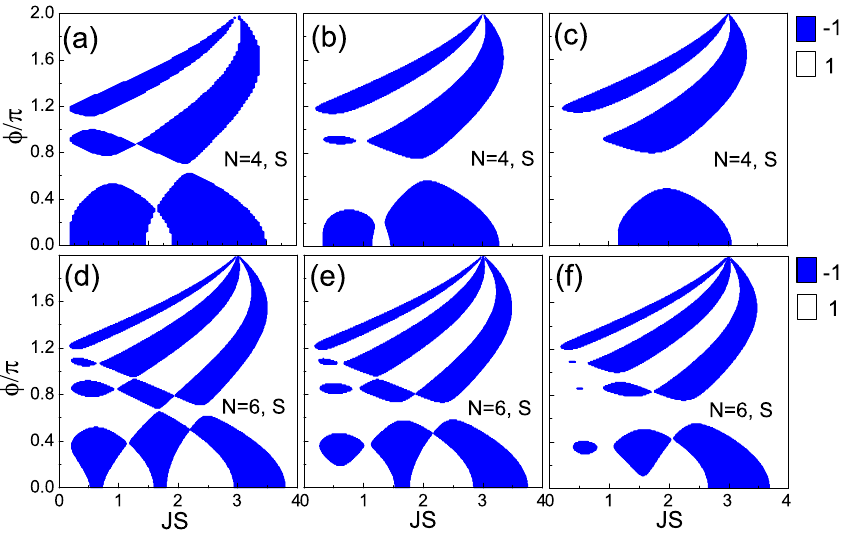}}
\caption{\label{fig:epsart} The $\mathbb{Z}_{2}$-valued invariant as a function of the pitch angle $\phi$ and $JS$ for the noncoplananr single-cone S states without the chiral symmetry. The parameters for the upper three panels are $N=4$, $\theta_{1}=\pi-\theta_{2}=0.4\pi$ while $\theta_{0}=\pi-\theta_{3}=0.3\pi$ for (a), $0.1\pi$ for (b), and $0$ for (c). The parameters for the lower three panels are $N=6$, $\theta_{1}=\pi-\theta_{4}=0.3\pi$, $\theta_{2}=\pi-\theta_{3}=0.4\pi$, while $\theta_{0}=\pi-\theta_{5}=0.2\pi$ for (d), $0.1\pi$ for (e), $0$ for (f). Here $t'=0.8t$, $\mu=-1$.}
\end{figure}

Based on the discussions above, we summarize in table I the symmetries and topological classifications of the five noncollinear phases.

\begin{table*}[!htp]
\caption{Topological properties of the five self-consistently determined uniform noncollinear magnetic phases in coupled triangular $N$-chain ladders. The presence (absence) of chiral symmetry is denoted by ``\checkmark'' (``$\times$''). The chiral operator $S$ is written in Nambu basis where the particle-hole symmetry operator $C$ takes the form $C=\tau_{y}\sigma_{y}K$, with $K$ the complex conjugation. Here $\bm{\tau}$ and $\bm{\sigma}$ are particle-hole and spin Pauli matrices respectively. The $\Gamma$ matrix acts as the reflection operator in the interchain index space.}
\begin{tabular}{c|cccc|c}
\hline
\hline
Magnetic phases& CI & S & DI & DII & \\
\hline

\multirow{2}{*}{Chiral symmetry} & \checkmark & $\times$ & $\diagdown$ & $\diagdown$ & N even\\

 & \checkmark & \checkmark & \checkmark & \checkmark if $\theta_{\frac{N-1}{2}}=\frac{\pi}{2}$ ($\times$ otherwise) &N odd\\

\multirow{2}{*}{Chiral operator $S$} &\multirow{2}{*}{$\tau_{y}\sigma_{z}$ or $\Gamma\tau_{y}\sigma_{z}$}& $\diagdown$ & $\diagdown$ & $\diagdown$ & N even\\

& & $\Gamma\tau_{y}\sigma_{z}$ & $\Gamma\tau_{y}\sigma_{z}$ & $\Gamma\tau_{y}\sigma_{z}$ if $\theta_{\frac{N-1}{2}}=\frac{\pi}{2}$ ($\diagdown$ otherwise) & N odd\\

\multirow{3}{*}{Topological classification} & \multirow{3}{*}{BDI[$\mathbb{Z}$]} & D[$\mathbb{Z}_{2}$] & $\diagdown$ & $\diagdown$ & N even\\

& & \multirow{2}{*}{BDI[$\mathbb{Z}$]} & \multirow{2}{*}{BDI[$\mathbb{Z}$]} & BDI[$\mathbb{Z}$] if $\theta_{\frac{N-1}{2}}=\frac{\pi}{2}$ & \multirow{2}{*}{N odd}\\
&&&&(D [$\mathbb{Z}_{2}$] otherwise)&\\
\hline
\end{tabular}
\end{table*}
\subsection{Extension to coupled square ladders}

\subsubsection{Magnetic structures.}

Our main results can be straightforwardly extended to coupled $N$-chain square ladders. For $N$-chain square ladder, certain coplanar magnetic structures and their topological properties have been studied\cite{Poyhonen2014}. The magnetic structures for the ground states are still believed to be combinations of the three basic structures but may show more complex situations. Here parallel to the discussions of the triangular ladder, we restrict our discussions to the following three noncollinear phases:
\begin{enumerate}
\item Coplanar phases(C), $S_{j}(l)=S(\cos[j\phi+l\phi'],\sin[j\phi+l\phi'],0)$;
\item Single-cone phases(S), $S_{j}(l)=S(\sin\theta_{l}\cos[j\phi+l\phi'],\sin\theta_{l}\sin[j\phi+l\phi'],\cos\theta_{l})$, with $\theta_{l}=\pi-\theta_{N-1-l}$;
\item Double-cone phases(D), $S_{j}(l)=S(\sin\theta_{l}\cos[j\phi+l\phi'],\sin\theta_{l}\sin[j\phi+l\phi'],(-1)^{j}\cos\theta_{l})$, with $\theta_{l}=\pi-\theta_{N-1-l}$,
\end{enumerate}
where $\phi'$ can only be $0$ or $\pi$. Our SC calculations for a coupled double chain with $t'=0.8$ and $JS=0.2$ indicate that the coplanar C states with $\phi'=0$ are stabilized when electron density $n$ is less than $0.24$, a collinear state is favored when $0.24<n<0.4$, while the coplanar C states with $\phi'=\pi$ are stabilized when $0.4<n<1$.

\subsubsection{Topological classification and topological invariants.}

\begin{figure*}
\scalebox{1.0}{\includegraphics[width=0.95\textwidth]{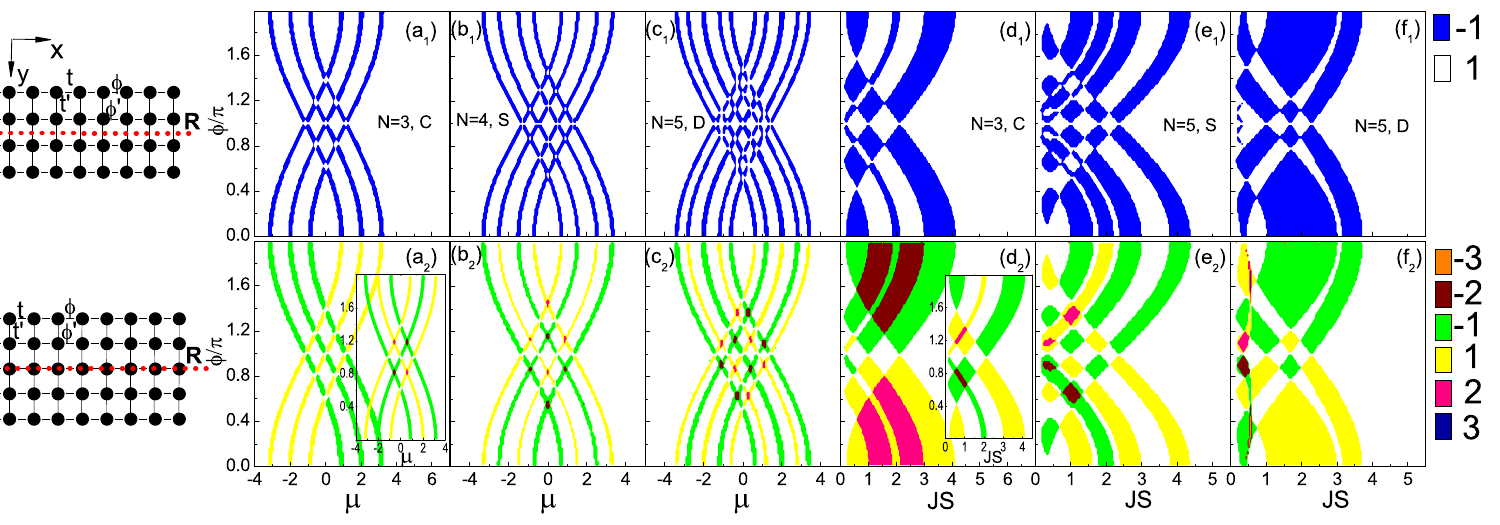}}
\caption{\label{fig:epsart} The $\mathbb{Z}_{2}$-valued invariants $(\mathrm{a}_{1})-(\mathrm{f}_{1})$ and the corresponding $\mathbb{Z}$-valued invariants $\nu$ $(\mathrm{a}_{2})-(\mathrm{f}_{2})$ for coupled $N$-chain square ladders with $\phi'=0$. Here $(\mathrm{a}_{1})$, $(\mathrm{a}_{2})$, $(\mathrm{d}_{1})$ and $(\mathrm{d}_{2})$ are for the coplanar states with $N=3$, while the others are for the noncoplanar ones. The topologically nontrivial states exhibited in $(\mathrm{a}_{2})$, $(\mathrm{d}_{2})$ are protected by the chiral symmetry relevant to $\tau_{y}\sigma_{z}$ while the insets in them by that relevant to $\Gamma\tau_{y}\sigma_{z}$. The parameters for the others are: $\theta_{0}=\pi-\theta_{3}=0.3\pi$, $\theta_{1}=\pi-\theta_{2}=0.4\pi$ for $(\mathrm{b}_{1})$,$(\mathrm{b}_{2})$; $\theta_{0}=\pi-\theta_{4}=0.3\pi$, $\theta_{1}=\pi-\theta_{3}=0.4\pi$, $\theta_{2}=0.5\pi$ for $(\mathrm{e}_{1})$,$(\mathrm{e}_{2})$; $\theta_{1}=\pi-\theta_{3}=0.3\pi, \theta_{0}=\theta_{2}=\theta_{4}=0.5\pi$ for $(\mathrm{c}_{1})$,$(\mathrm{c}_{2})$, $(\mathrm{f}_{1})$,$(\mathrm{f}_{2})$. Here $t'=0.8$, and $JS=0.2$ is fixed for the left six panels, while $\mu=-1$ is fixed for the right six panels. }
\end{figure*}

In the exactly similar basis to the triangular ladder cases, the $k$-space Hamiltonian for the above three kind of states can be written as,
\begin{equation}
\widetilde{H}_{k}^{C}=H_{k}^{0}+\widetilde{\xi}\Gamma_{x}\tau_{z}+\widetilde{\eta}\Gamma_{y}\tau_{z}\sigma_{z},
\end{equation}
for C states, with $\widetilde{\xi}=-t'\cos\frac{\phi'}{2}$, $\widetilde{\eta}=-t'\sin\frac{\phi'}{2}$;
\begin{equation}
\widetilde{H}_{k}^{S}=\widetilde{H}_{k}^{'0}+h\Gamma_{z}^{1}\sigma_{x}+h\Gamma_{z}^{2}\sigma_{z}+\widetilde{\xi}\Gamma_{x}\tau_{z}+\widetilde{\eta}\Gamma_{y}\tau_{z}\sigma_{z},
\end{equation}
for single-cone S states, where $\widetilde{H}_{k}^{'0}=(\xi_{k,\phi}-\mu)\tau_{z}+\eta_{k,\phi}\tau_{z}\sigma_{z}+\Delta\tau_{x}$;
\begin{equation}
\widetilde{H}_{k}^{D}=\widetilde{H}_{k}^{''0}+h\Gamma_{z}^{1}\sigma_{x}+hr_{z}\Gamma_{z}^{2}\sigma_{z}+\widetilde{\xi}\Gamma_{x}\tau_{z}+\widetilde{\eta}\Gamma_{y}\tau_{z}\sigma_{z},
\end{equation}
for the double-cone D states, where $\mathcal{H}_{k}^{''0}=(\cos\frac{k}{2}r_{x}+\sin\frac{k}{2}r_{y})\tau_{z}(\xi_{k/2,\phi}+\eta_{k/2,\phi}\sigma_{z})-\mu\tau_{z}+\Delta\tau_{x}$, with $r_{i}$ the sublattice Pauli matrices. For the coplanar C states with $\phi'=0$, the signature $M$ of the Pfaffian can be given analytically,
\begin{equation}
M=\mathrm{sgn}\prod_{n=1}^{N}(\delta^{2}-\xi_{n+}^{2})(\delta^{2}-\xi_{n-}^{2}),
\end{equation}
where $\xi_{n\pm}=\pm2t\cos\frac{\phi}{2}-2t^{'}\cos(\frac{n\pi}{N+1})-\mu$. While for $\phi'=\pi$, $M$ is trivial for even $N$, but $M=\mathrm{sgn}\{[\delta^{2}-(2\cos\frac{\phi}{2}-\mu)^{2}][\delta^{2}-(-2\cos\frac{\phi}{2}-\mu)^{2}]\}$ for odd $N$.

\begin{figure}
\centering
\scalebox{1.0}{\includegraphics[width=0.45\textwidth]{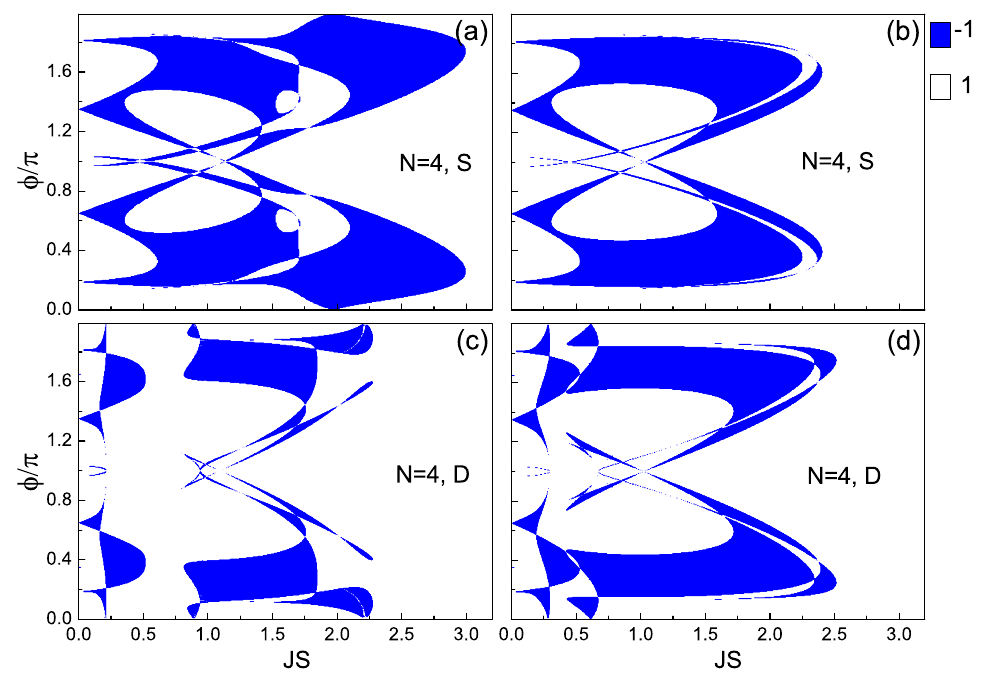}}
\caption{\label{fig:epsart} The $\mathbb{Z}_{2}$ invariant $N_{1D}$ for the noncoplanar S and D phases of class DIII in coupled 4-chain square ladders. The topologically nontrivial regions are denoted by blue color ($N_{1D}=-1$). Here $\phi'=\pi$, $t'=0.8$, $\mu=-1$. The other parameters are: $\theta_{0}=\pi-\theta_{3}=0.3\pi$, $\theta_{1}=\pi-\theta_{2}=0.4\pi$ for (a)(c); $\theta_{0}=\pi-\theta_{3}=0.4\pi$, $\theta_{1}=\pi-\theta_{2}=0.45\pi$ for (b)(d). }
\end{figure}

One can check that the chiral symmetry is preserved for all the phases mentioned above as shown in table II. The fact that the chiral symmetry for a coupled square ladder is always preserved for the noncoplanr phases, regardless of the parity of $N$, is because there always exists the reflection line, as shown in the insets of figure 10. Concretely, the chiral operator $S$ satisfying $\{H_{k}^{C(S,D)},S\}=0$ takes the form: $S=\tau_{y}\sigma_{z}$ for the planar C states with $\phi'=0$ or $\pi$; while $S=\Gamma\tau_{y}\sigma_{z}$  for the noncoplanar S or D states with $\phi'=0$, and $S=\widetilde{\Gamma}\tau_{y}\sigma_{z}$ for the noncoplanar S or D states with $\phi'=\pi$, with
\begin{equation}
\begin{array}{c}
\widetilde{\Gamma}=
\end{array}
\left(
\begin{array}{ccccc}
  &   &   &   &    -1\\
  &   &   & 1 &  \\
  &   &  -1 &   &  \\
  & \cdots &   &   &  \\
(-1)^{N} &   &   &   &  \\
\end{array}
\right),
\end{equation}
since it can be easily shown that $\{\widetilde{\Gamma},\Gamma_{x}\}=[\widetilde{\Gamma},\Gamma_{y}]=0$ and the same commutation or anticommutation relations with $\Gamma_{z}^{1}$, $\Gamma_{z}^{2}$ to $\Gamma$: $[\widetilde{\Gamma},\Gamma_{z}^{1}]=\{\widetilde{\Gamma},\Gamma_{z}^{2}\}=0$. $\widetilde{\Gamma}$ can be regarded as the `new' reflection operator in the interchain index space for $\phi'=\pi$ case. The existence of the chiral symmetries, together with the particle-hole symmetry lead naturally to $T$ symmetries: $T\widetilde{H}_{-k}T^{-1}=\widetilde{H}_{k}$, with $T=\sigma_{x}K$ for the coplanar states, $T=\Gamma\sigma_{x}K$ for the noncoplanar ones with $\phi'=0$, and $T=\widetilde{\Gamma}\sigma_{x}K$ for the noncoplanar ones with $\phi'=\pi$. Therefore, although all the above states are chiral symmetric, they belong to two different topological classes. All the states belong to BDI class,
except the noncoplanar states with $\phi'=\pi$ and even $N$, which belong to DIII class, since $T^{2}=-1$ for even $N$\cite{Andreas2008,Ryu2010}. DIII class is characterized by a $\mathbb{Z}_{2}$-valued invariant, which is the parity of the Kramer's pairs of Majorana fermions at each end. In the weak pairing limit,
the $\mathbb{Z}_{2}$ invariant $N_{1D}$ for a 1D class DIII superconductor can be expressed as follows\cite{Qi2010},
\begin{equation}
N_{1D}=\prod_{\alpha}sgn(g_{\alpha}),
\end{equation}
where $\alpha$ is summed over all the Fermi points between 0 and $\pi$, and $g_{\alpha}$ = $<\alpha{k_{F}}|i\widetilde{\Gamma}\sigma_{z}|\alpha{k_{F}}>$, with $|\alpha{k}>$ the Bloch states of band $\alpha$ for the normal-state Hamiltonian of $\widetilde{H}_{k}^S$ or $\widetilde{H}_{k}^D$. The corresponding topological phase diagrams as functions of the pitch angle $\phi$ and $JS$ are shown in figure 11.

\begin{figure}
\centering
\scalebox{1.0}{\includegraphics[width=0.45\textwidth]{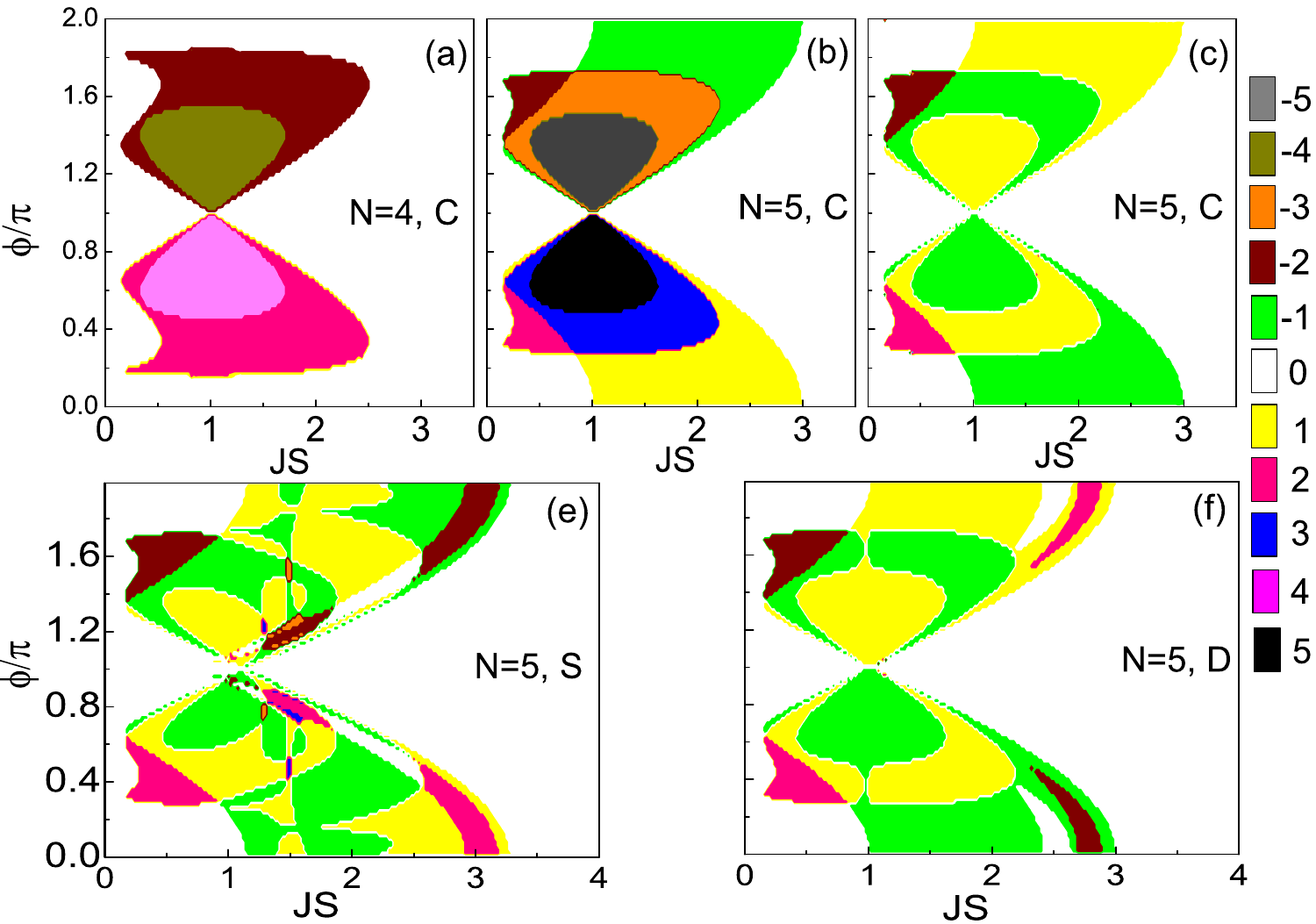}}
\caption{\label{fig:epsart} The $\mathbb{Z}$-valued invariants for coupled $N$-chain square ladders with $\phi'=\pi$, where (a),(b) and (c) are for the planar states while (e),(f) are for the single-cone S and double-cone D states respectively. The integer values shown in (a)(b) corresponds to the chiral symmetry relevant to $\tau_{y}\sigma_{z}$ while (c) to that relevant to $\widetilde{\Gamma}\tau_{y}\sigma_{z}$. The parameters for the latter two are: (e)$N=5$, $\theta_{0}=\pi-\theta_{4}=0.3\pi$, $\theta_{1}=\pi-\theta_{3}=0.4\pi$, $\theta_{2}=0.5\pi$; (f)$N=5$, $\theta_{1}=\pi-\theta_{3}=0.4\pi, \theta_{0}=\theta_{2}=\theta_{4}=0.5\pi$. Here $t'=0.8$, $\mu=-1$.}
\end{figure}
For the planar phases, the $\mathbb{Z}$ invariants $\nu$ can be expressed as the sum of that of the independent decoupled chains, $\nu=\sum_{n=1}^{N}\nu_{n}$, since $[\widetilde{H}_{k}^{C},\Gamma_{x(y)}]=0$ when $\phi'=0(\pi)$ and thus $\widetilde{H}_{k}^{C}$ can be reduced and then classified by the eigenvalues of $\Gamma_{x(y)}$. Here $\nu_{n}$ for $\phi'=0 $ can be derived to be $\nu_{n}=\mathrm{sgn}(\delta^{2}-\xi_{n+}^{2})$ if $(\delta^{2}-\xi_{n+}^{2})(\delta^{2}-\xi_{n-}^{2})<0$ and $\nu_{n}=0$ otherwise. The $\mathbb{Z}_{2}$-valued invariant $M$ and the $\mathbb{Z}$-valued invariant $\nu$ for coupled square ladders are depicted in figure 10. Compared with the triangular cases, the $\phi$ versus $\mu$ phase diagrams for $M$ generically consist of $2N$ branches and show the symmetric structures: $M(-\phi,\mu)=M(\phi,-\mu)=M(\phi,\mu)$, while $\nu(-\phi,\mu)=\nu(\phi,-\mu)=-\nu(\phi,\mu)$. The $\phi$ versus $JS$ phase diagram also exhibits the similar symmetric structures with respect to $\phi$, but regions with high odd integers such as $\nu=\pm3$ could occur at larger values of $JS$.

We notice here that similar to the triangular ladder case, the planar phase with $\phi'=0$($\pi$) actually has two chiral operators. One is $S=\tau_{y}\sigma_{z}$, the other is $S'=\Gamma\tau_{y}\sigma_{z}(\widetilde{\Gamma}\tau_{y}\sigma_{z})$ since $[\widetilde{H}_{k}^{C},\Gamma(\widetilde{\Gamma})]=0$ for $\phi'=0$($\pi$). Accordingly, there is another topological invariant $\nu'$ relevant to $S'$. The $\nu'$ for $\phi'=0$ can also be given analytically: $\nu'=\sum_{n=1}^{N}(-1)^{n}\nu_{n}$, where $\nu_{n}$ takes exactly the same form to that for $\phi'=0$. The factor $(-1)^{n}$ can be interpreted as the eigenvalue of $\Gamma$ for the $n$th eigenstate of $\Gamma_{x}$ or $\Gamma_{y}$. The $\mathbb{Z}$ invariants $\nu$ and $\nu'$ are compared with each other in figure 10($\mathrm{a}_{2}$),($\mathrm{d}_{2}$) and their insets. What's different from the triangular case is that relevant to $S=\tau_{y}\sigma_{z}$ the coplanar phase with $\phi'=\pi$ for even $N$ belongs to BDI class and can thus be characterized by $\mathbb{Z}$ invariants, while relevant to $S'=\widetilde{\Gamma}\tau_{y}\sigma_{z}$ the same phase belongs to DIII class and can thus be characterized by $\mathbb{Z}_{2}$ invariants. This novel feature can be understood if we note that the MBSs at ends in this case can be protected by one of the chiral symmetries or by both. Under the reflection-symmetry-breaking perturbations, for the former situation, the MBSs protected by $S'$ can be removed, while for the latter, the MBSs protected by $S'$ can still survive since they are also protected by the chiral symmetry relevant to $S$.

\begin{table*}[!htp]

\caption{Topological properties of the three uniform noncollinear magnetic phases in coupled square $N$-chain ladders with $\phi^{'}$= 0 and $\phi^{'}$= $\pi$ respectively. The presence (absence) of chiral symmetries is denoted by ``\checkmark'' (``$\times$''). The chiral operator $S$ is written in Nambu basis where the particle-hole symmetry operator $C$ takes the form $C=\tau_{y}\sigma_{y}K$, with $K$ the complex conjugation. Here $\bm{\tau}$ and $\bm{\sigma}$ are particle-hole and spin Pauli matrices respectively. The $\Gamma$($\widetilde{\Gamma}$) matrix acts as the reflection operator in the interchain index space for $\phi^{'}$= 0($\phi^{'}$= $\pi$).}
\begin{tabular}{c|ccc|ccc|c}
\hline
\hline
\multirow{2}{*}{}&\multicolumn{3}{c|}{$\phi^{'}=0$} & \multicolumn{3}{c|}{$\phi^{'}=\pi$}&\\
\hline

Magnetic phases & C & S & D & C & S & D &\\

Chiral symmetry& \checkmark & \checkmark & \checkmark & \checkmark & \checkmark & \checkmark & \\

Chiral operator $S$ & $\tau_{y}\sigma_{z}$ or $\Gamma\tau_{y}\sigma_{z}$& $\Gamma\tau_{y}\sigma_{z}$ & $\Gamma\tau_{y}\sigma_{z}$ & $\tau_{y}\sigma_{z}$ or $\widetilde{\Gamma}\tau_{y}\sigma_{z}$& $\widetilde{\Gamma}\tau_{y}\sigma_{z}$ & $\widetilde{\Gamma}\tau_{y}\sigma_{z}$ & \\

\multirow{2}{*}{Topological classification} & BDI[$\mathbb{Z}$] & BDI[$\mathbb{Z}$] & BDI[$\mathbb{Z}$] & BDI[$\mathbb{Z}$] or DIII[$\mathbb{Z}_{2}$]& DIII[$\mathbb{Z}_{2}$]  & DIII[$\mathbb{Z}_{2}$] &N even\\

& BDI[$\mathbb{Z}$] & BDI[$\mathbb{Z}$] & BDI[$\mathbb{Z}$] & BDI[$\mathbb{Z}$]&BDI[$\mathbb{Z}$]  & BDI[$\mathbb{Z}$] &N odd\\
\hline
\end{tabular}

\end{table*}

The $\mathbb{Z}$-valued invariants for the chiral symmetric states with $\phi'=\pi$ in classification of BDI class are shown in figure 12, which exhibits the symmetric butterfly-like structures. The nontrivial region of the signature $M$ can be regarded as that of $\nu$ with odd integer values. We note that the $\nu$ values for the noncoplanar states with even $N$ in this case are inevitably zero due to $T$ symmetry. For the noncoplanar phases, no matter in BDI or DIII class, further discussions lead to the conclusion that the nonuniform $\Delta$ or $\mu$ terms with the reflection symmetry still respect the chiral symmetry and thus cannot change the number of MBSs at ends. The symmetries and topological classifications for the above mentioned phases are summarized in table II.

\section{\label{sec:level4}Summary}
In summary, we have studied the possibility of the topological superconductivity for coupled triangular magnetic atomic chains deposited on a conventional superconductor. A variety of noncollinear magnetic structures have been determined by self-consistently solving the BdG equations, including the coplanar and noncoplanar ones, which can be regarded as combinations of three basic single-chain structures: helix, single-cone helix, and double-cone helix. All these self-organized structures support topologically nontrivial superconducting states which can be characterized by $\mathbb{Z}_{2}$-valued invariants due to particle-hole symmetry. If the magnetic structure is coplanar, or noncoplanar with an odd chain number, the systems have chiral symmetries and thus belong to BDI class and so can be characterized by $\mathbb{Z}$-valued invariants, which represent the numbers of Majorana bound states at ends. Although the coplanar and noncoplanar states respect different chiral symmetries, these chiral symmetries do not rely on the translational symmetry along the chain and are also robust in the presence of weak on-site disorder as long as the reflection symmetry is respected. The parallel discussions and similar conclusions are also made for coupled square atomic ladders. However, for square ladders with even chain number there exist the noncoplanar phases of class DIII characterized by $\mathbb{Z}_{2}$-valued invariants.

Experimentally, magnetic spiral structures have been detected by spin-polarized STM in a single atomic layer Mn on a W(110) substrate \cite{Bode2007} and in bi-atomic Fe chains on a reconstructed Ir(001) surface\cite{Menzel2012}. Moreover, magnetic atoms' spins can be manipulated atom by atom by STM\cite{Serrate2010,Khajetoorians2011,Khajetoorians2012}. For a large doping range, our self-consistent results show the magnetic moments of atoms can be so self-organized to be helical due to RKKY interactions that the system is topologically superconducting. This indicates that by tuning the gate voltage on the superconductor, arrays of magnetic atoms on it may be expected to realize topological superconductivity, without the need for pinning the magnetic moments of atoms by STM.

\begin{acknowledgments}
This work was supported by NSFC under grant No.11174126, and the State Key Program for Basic Researches of China under grant No.2015CB921202.
\end{acknowledgments}


\end{document}